\documentclass[prb,twocolumn,nopacs,superscriptaddress,floatfix]{revtex4}

\usepackage{graphicx}
\usepackage[centertags]{amsmath}
\usepackage{color}

\definecolor{grey}{rgb}{.65,.65,.65}

\newcommand{\cm}{\ensuremath{\,\mbox{cm}^{-1}}}
\newcommand{\K}{\ensuremath{\,\mbox{K}}}
\newcommand{\celsius}{\ensuremath{^\circ}C}

\newcommand{\Tc}{\ensuremath{T_{\rm C}}}
\newcommand{\Tn}{\ensuremath{T_{\rm N}}}

\begin{document}

\title{Unusual ferroelectric and magnetic phases in multiferroic 2H-BaMnO$_3$ ceramics}

\author{Stanislav Kamba}
\thanks{Corresponding author: kamba@fzu.cz}{}
\affiliation{Institute of Physics, Czech Academy of Sciences, Na Slovance~2, 182 21 Prague~8, Czech Republic}
\author{Dmitry Nuzhnyy}
\affiliation{Institute of Physics, Czech Academy of Sciences, Na Slovance~2, 182 21 Prague~8, Czech Republic}
\author{Maxim Savinov}
\affiliation{Institute of Physics, Czech Academy of Sciences, Na Slovance~2, 182 21 Prague~8, Czech Republic}
\author{Pierre Tol\'{e}dano}
\affiliation{Laboratoire de Physique de Syst\`{e}mes complexes, Universit\'e de Picardie, 80000 Amiens, France}
\author{Valentin Laguta}
\affiliation{Institute of Physics, Czech Academy of Sciences, Na Slovance~2, 182 21 Prague~8, Czech Republic}
\author{Petr Br\'{a}zda}
\affiliation{Institute of Physics, Czech Academy of Sciences, Na Slovance~2, 182 21 Prague~8, Czech Republic}
\author{Luk\'{a}\v{s} Palatinus}
\affiliation{Institute of Physics, Czech Academy of Sciences, Na Slovance~2, 182 21 Prague~8, Czech Republic}
\author{Filip Kadlec}
\affiliation{Institute of Physics, Czech Academy of Sciences, Na Slovance~2, 182 21 Prague~8, Czech Republic}
\author{Fedir Borodavka}
\affiliation{Institute of Physics, Czech Academy of Sciences, Na Slovance~2, 182 21 Prague~8, Czech Republic}
\author{Christelle Kadlec}
\affiliation{Institute of Physics, Czech Academy of Sciences, Na Slovance~2, 182 21 Prague~8, Czech Republic}
\author{Petr Bednyakov}
\affiliation{Institute of Physics, Czech Academy of Sciences, Na Slovance~2, 182 21 Prague~8, Czech Republic}
\author{Viktor Bovtun}
\affiliation{Institute of Physics, Czech Academy of Sciences, Na Slovance~2, 182 21 Prague~8, Czech Republic}
\author{Martin Kempa}
\affiliation{Institute of Physics, Czech Academy of Sciences, Na Slovance~2, 182 21 Prague~8, Czech Republic}
\author{Dominik Kriegner}
\affiliation{Faculty of Mathematics and Physics, Charles University, Ke Karlovu~5, 121 16 Prague~2, Czech Republic}
\author{Jan Drahokoupil}
\affiliation{Institute of Physics, Czech Academy of Sciences, Na Slovance~2, 182 21 Prague~8, Czech Republic}
\author{Jan Kroupa}
\affiliation{Institute of Physics, Czech Academy of Sciences, Na Slovance~2, 182 21 Prague~8, Czech Republic}
\author{Jan Prokle\v{s}ka}
\affiliation{Faculty of Mathematics and Physics, Charles University, Ke Karlovu~5, 121 16 Prague~2, Czech Republic}
\author{Kamal Chapagain}
\affiliation{Department of Physics, Northern Illinois University, DeKalb, IL, USA}
\author{Bogdan Dabrowski}
\affiliation{Department of Physics, Northern Illinois University, DeKalb, IL, USA}
\author{Veronica Goian}
\affiliation{Institute of Physics, Czech Academy of Sciences, Na Slovance~2, 182 21 Prague~8, Czech Republic}

\begin{abstract}

The structural phase transition in hexagonal BaMnO$_3$ occurring at
\Tc=130\,K was studied in ceramic samples using electron and X-ray diffraction, second harmonic generation as well as by
dielectric and lattice dynamic spectroscopies. The low-temperature phase
(space group $P6_{3}cm$) is ferroelectric with a triplicated unit
cell. The phase transition is driven by an optical soft
mode from the Brillouin-zone boundary [$q = (\frac{1}{3},\frac{1}{3},0)$];
this mode activates in infrared and Raman spectra below \Tc{} and
it hardens according to the Cochran law. Upon cooling below
\Tc, the permittivity exhibits an unusual linear increase with temperature;
below 60\,K, in turn, a frequency-dependent decrease is observed, which can be
explained by slowing-down of ferroelectric domain wall motions. Based on our data we could
not distinguish whether the high-temperature phase is paraelectric or
polar (space groups $P6_{3}/mmc$ or $P6_{3}mc$, respectively). Both variants of
the phase transition  to
the ferroelectric phase are
discussed based on the Landau theory. Electron paramagnetic resonance
and magnetic susceptibility measurements reveal an onset
of one-dimensional antiferromagnetic ordering below
$\approx220\,\rm K$ which develops fully near 140\,K and, below $\Tn
\approx 59\,\rm K$, it transforms into a three-dimensional antiferromagnetic order.

\end{abstract}
\date{\today}

\maketitle

\section{Introduction}

In proper ferroelectrics, the polarization plays the role of the order parameter
(OP) and at the Curie temperature \Tc, a large peak in the temperature
dependence of the permittivity $\varepsilon^\prime(T)$ is observed due to softening
of some polar excitation.\cite{Scott74,Petzelt16}  In displacive proper ferroelectrics,
this excitation corresponds to a polar phonon active in far-infrared (IR) spectra in
both paraelectric and ferroelectric (FE) phases. At \Tc, $\varepsilon^\prime(T)$ exhibits a
maximum due to the phonon softening, since the soft-phonon frequency $\omega_{\rm SM}$ follows
the Cochran law, $\omega_{\rm SM}(T)=A\sqrt{T-\Tc}$, and according to the
Lyddane-Sachs-Teller relation, the static permittivity
$\varepsilon^\prime(T) \propto  1/\omega_{\rm SM}^{2}$($T$).\cite{Scott74,Petzelt16} In proper
ferroelectrics with order-disorder-type phase transitions, phonons are stable;
by contrast, these compounds exhibit a dielectric relaxation whose frequency $\omega_{\rm r}$
is located in the MHz--GHz region.\cite{Grigas96} On cooling towards \Tc, this
frequency decreases according to a modified Cochran law,
$\omega_{\rm r}=A^\prime(T-\Tc)$, which induces an anomaly in $\varepsilon^\prime(T)$ at
\Tc. As the value of $\varepsilon^\prime$ near \Tc{} is strongly frequency-dependent,
a maximum or minimum in
  $\varepsilon^\prime(T)$ at \Tc{} is observed if the measuring frequency is lower or
  higher than $\omega_{\rm r}$, respectively.\cite{Grigas96} It is worth noting that many proper
ferroelectrics exhibit phase transitions corresponding to a crossover from
displacive to order-disorder
type\cite{Petzelt87,Buixaderas04,Hlinka08,Kadlec11}. In such a case, some polar phonon softens on
cooling far above \Tc, but this softening ceases at some temperature closer to
\Tc{} as an additional soft dielectric relaxation (called a central mode) appears.\cite{Scott74,Petzelt87,Buixaderas04}

In pseudoproper and improper ferroelectrics, the OP is represented by another
quantity such as strain, the antiferromagnetic (AFM) OP, the eigenvector of a phonon with a
wavevector off the Brillouin zone (BZ)
center,\cite{Levanyuk74,Dvorak74,Toledano09} charge or orbital
ordering.\cite{Young15} In these cases the FE
polarization below \Tc{} arises due to coupling of the polarization with the
primary OP and usually only a small dielectric anomaly appears near \Tc{}.
Proper and pseudoproper ferroelectrics exhibit equitranslational phase
transitions, whereas improper FE phase transitions are
non-equitranslational, i.e., below \Tc{}, a multiplication of the unit cell
occurs. The pseudoproper ferroelectrics include most of the spin-induced
ferroelectrics (type-II multiferroics) featuring an effective bilinear coupling
of the polarization with the AFM OP.\cite{Toledano09}
Applications of anharmonic lattice interactions in improper ferroelectrics for
the design of multiferroics are summarized in a recent review.\cite{Young15}

Type-I multiferroics are mostly proper ferroelectrics---their polarization exists
independently of their magnetization. The exceptions include hexagonal RMnO$_3$
materials ($\rm R = Sc$, Y, Dy--Lu) which exhibit improper FE phase transitions
at 900--1300\,K \cite{Lilienblum15} connected with a tripling of the unit cell
below \Tc{}, and AFM phase transitions around
70--130\,K.\cite{Munoz00,Yen07} Their structural phase transitions were
theoretically predicted to be induced by a $K_{3}$-symmetry soft mode from the
BZ boundary [$q = (\frac{1}{3},\frac{1}{3},0)$]\cite{Fennie05}; however, this
was never confirmed, because the required inelastic neutron or X-ray scattering
experiments above $\Tc\approx1000\,\rm K$ are quite difficult to perform.
Moreover, the phonons at high temperatures are always heavily damped, which makes
their detection even more difficult. The soft mode should become IR and Raman
active below \Tc, but no such mode was revealed in recent IR\cite{Goian10a} or Raman
studies.\cite{Bouyanfif15}

The lattice dynamics were studied also in other improper ferroelectrics, such as
$\rm Gd_2(MoO_4)_3$\cite{Petzelt84} or boracites\cite{Lockwood76,Petzelt73}, but
no soft mode was detected\cite{Lockwood76,Petzelt73} or it was
markedly overdamped and therefore it was seen as a central
mode.\cite{Petzelt84} Recently, Varignon and Ghosez\cite{Varignon13} predicted
an improper FE phase transition in the two-layered hexagonal BaMnO$_3$
(further denoted as 2H-BaMnO$_3$).
Interestingly, in analogy with RMnO$_3$, the phase transition should be driven
by a $K_{3}$-symmetry soft mode with $q = (\frac{1}{3},\frac{1}{3},0)$ and the
polarization should be induced by a coupling between the soft mode and a hard
zone-center mode of $\Gamma_{2}^{-}$ symmetry.

Although most ABO$_3$ compounds  crystallize in the
perovskite structure, this does not hold true for BaMnO$_3$ whose
perovskite polymorph is unstable; instead, it crystallizes in a more
stable hexagonal form with layers of face-sharing MnO$_6$ octahedra.
The arrangement of these layers is  determined by the synthesis conditions
(temperature and atmosphere) which may produce oxygen vacancies. As their
concentration increases, the double-layered structure (i.e.,\ $Z=2$) gradually
transforms to 15, 8, 6, 10 and 4-layered ones.\cite{Negas71,Adkin07} In this
paper we deal with the stoichiometric
2H-BaMnO$_3$, whose room-temperature crystal structure has raised a
long-standing discussion. Some authors reported a non-polar space group
$P6_{3}/mmc$ ($Z=2$)\cite{Christensen72,Cussen00,Varignon13} but other
structural and vibrational studies proposed a polar one, $P6_{3}mc$
 ($Z=2$).\cite{Roy98,Stanislavchuk15} Also the low-temperature
crystal structure was uncertain for a long time. Cussen and Battle, who had
refined the crystal structure at 80\,K, concluded that the space group
is $P6_{3}cm$ and $Z=6$. Only very recently Stanislavchuk \textit{et al.}\
confirmed this structure by neutron diffraction, finding the structural
phase transition at $\Tc{}=130\pm5\,\rm K$.\cite{Stanislavchuk15} They
investigated also the lattice dynamics of 2H-BaMnO$_3$ using Raman scattering
and far-IR ellipsometry, detecting much fewer modes than the number
  predicted by the factor-group analysis, and no soft mode.  At low
  temperatures, 2H-BaMnO$_3$ exhibits an AFM order, but the  details are
    also controversial; according to Christensen and Ollivier,
    \cite{Christensen72} below the N\'{e}el temperature of $\Tn=2.4\,K$, the
    spins would be directed along the \textit{c} axis, while a newer work by
    Cussen and Battle\cite{Cussen00} reported $\Tn{} = 59\,K$ and the spin
    directions were identified in (001) planes.

Below we present results of a set of experiments on 2H-BaMnO$_3$ using
X-ray and electron diffraction, second harmonic generation (SHG), magnetic and EPR studies, dielectric, microwave, THz time-domain, Raman and IR spectroscopies. In the diffraction experiments, we confirmed the non-equitranslational phase
  transition to occur at $\Tc{}= 130\,$K. Furthermore, below \Tc, we observed an
  IR- and Raman-active soft phonon from the BZ boundary
which drives the phase transition. We show that dielectric permittivity and FE polarization exhibit
an unusual temperature behavior, which we discuss based on the
Landau theory. Above $\Tn\approx59\,\rm K$, magnetic susceptibility and EPR intensity do not
follow the Curie-Weiss behavior, which is related to a one-dimensional AFM order or
short-range magnetic correlations existing within at least 160\,K above \Tn.

\section{Experimental details}

\begin{figure*} \centering \includegraphics[width=180mm]{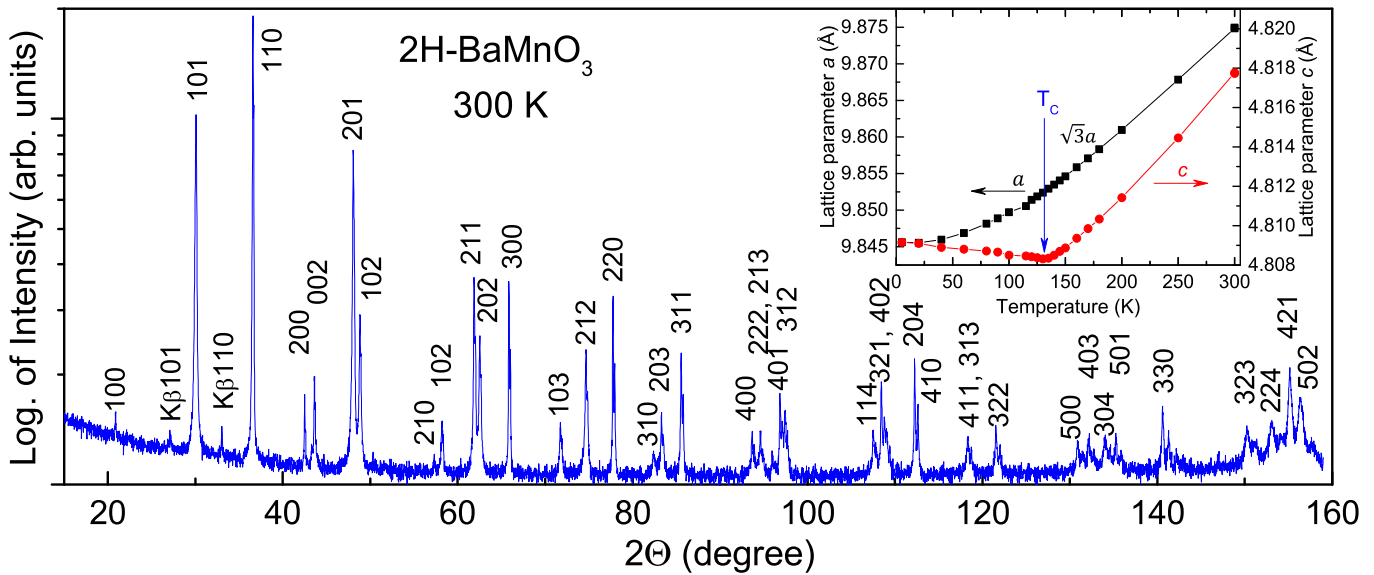}
  \caption{Room-temperature XRD pattern measured in the Bragg-Brentano geometry
    with a Co anode and the $K_{\alpha}$ line. A $\beta$ filter was used to
    avoid the $K_{\beta}$ radiation. Nevertheless, some minor diffraction peaks
    for the $K_{\beta}$ radiation were observed. All diffraction spots were
    assigned, and no extra peaks from other phases were detected. Inset shows
    the temperature dependences of lattice parameters obtained by XRD, which
    turn out to be the same as those obtained previously for single crystals of
    2H-BaMnO$_3$ in Ref.~\onlinecite{Stanislavchuk15} using neutron
  diffraction.} \label{fig:XRD-RT} \end{figure*}

 2H-BaMnO$_3$  was synthesized following the conventional ceramic
fabrication method from high-purity (99.999\,\%) BaCO$_3$ and MnO$_2$. The
starting mixture was wet-ball-milled for 12 hours in alcohol using an agate
container.  After drying, the resulting slush was calcined in oxygen at
800\,\celsius. The black powder was subsequently reground in an agate mortar and
pressed into high-density pellets. Several additional firings in oxygen with
intermediate grindings were carried out at increasing temperatures in the range
of 950--1150\,\celsius, followed by natural cooling to room temperature
for ca.\ 12
hours. To examine the phase purity of our samples, after each firing,
powder X-ray diffraction (XRD) measurements were carried out using a Rigaku D/MAX
powder diffractometer with CuK$_{\alpha}$ radiation in the $2\theta =
20$--$70^\circ$ range. In agreement with Ref.~\onlinecite{Negas71},  the single-phase 2H-BaMnO$_3$ formed only below 1125\,\celsius{} whereas above this
temperature, the 15-layered impurity phase was observed. The final ceramic samples were
thus prepared by firing in oxygen at 1100\,\celsius{} for 24
hours. Their structure was then again verified by XRD---see the diffraction pattern in
Fig.~\ref{fig:XRD-RT}, whereby a single-phase 2H-BaMnO$_3$ composition was
confirmed.

Low-temperature XRD was performed using a custom-adapted
Siemens D500 diffractometer equipped with a closed-cycle Sumitomo
Heavy Industries cryocooler enabling cooling down to 3\,K. We
used the Cu-$K\alpha_{1,2}$ radiation and the Bragg-Brentano geometry
with
source--sample and sample--detector distances of 330\,mm. The measurement was
performed with a fixed divergence slit size, resulting in
a primary-beam  divergence of 0.44$^\circ$. A MYTHEN 1K linear detector
and an optimized integration procedure,\cite{Kriegner15}  avoiding
geometrical defocussing, were used to obtain the powder diffraction
patterns. The lattice parameters were determined by Pawley mode in the Rietveld
analysis\cite{Rietveld69} using the Topas software tool.\cite{Cheary04}

Electron diffraction measurements were performed on a Philips CM120
  transmission electron microscope equipped with a Digistar
      Nanomegas precession device
and a Olympus Veleta CCD camera. The data were collected in
  the precession electron-diffraction tomography regime\cite{Mugnaioli09} with a
  precession angle of 0.6$^\circ$ and a tilt step of 0.6$^\circ$. A part of the pellet was
  crushed in an agate mortar and dispersed on a holey-carbon-coated Cu grid. The
  sample was inserted in a cooling holder with the low-temperature limit of 97\K.
  The data were processed with the  PETS software\cite{Palatinus11} and the structures
  refined with the Jana2006  software \cite{Petricek14}
  using the dynamical refinement method\cite{Palatinus15a,Palatinus15b}. Selected crystals with
lateral dimensions of less than 500\,nm (see Fig.~1 in the
Supplemental Material [\onlinecite{Suppl-2H-BaMnO3}]) were cooled down to 99\,K.
A full 3D electron diffraction data set (completeness 94\%) was collected at
temperatures from 99 to 240\,K. Finally, the crystal was cooled again to 120\,K to check for data consistency and possible radiation damage.

Radio-frequency dielectric measurements were performed using a NOVOCONTROL
Alpha-AN impedance analyzer in conjunction with a JANIS ST-100
cryostat (5--300\,K). Gold electrodes were evaporated on sample plates with a diameter of
3\,mm and thicknesses of 250 and 129\,$\mu$m. Dielectric hysteresis loops were measured
at a frequency of 50\,Hz with a testing voltage of 1\,V, using a custom-made
Sawyer-Tower bridge. Pyrocurrent measurements
(at a heating rate of 5\,K/min after a cooling with a bias of 40\,kV/cm) used a
KEITHLEY 6517A Electrometer high resistance meter; the resulting
polarization was calculated by subtracting the thermally stimulated current related to charged defects.

For microwave dielectric measurements, the method of a composite dielectric resonator
\cite{Bovtun08,Bovtun11} was used.  TE$_{01\delta}$ resonance modes were
excited first in the base cylindrical dielectric resonator and then in the composite
dielectric resonator, consisting of the base and the sample in the shielding
cavity. The disk-shaped sample had the same diameter ($\sim13$\,mm) as the base
and a thickness of 1.9\,mm. A comparison of the resonance frequencies and
quality factors of the base and composite resonators allowed for
calculating the dielectric parameters of the sample at the resonance frequency ($\sim6$\,GHz).
Temperature dependencies of the resonance frequencies and quality factors were
measured using an AGILENT E8364B vector network analyzer and a Janis closed-cycle He
cryostat (10\,K--400\,K).

SHG measurements were performed in order to confirm a possible
polar structure above \Tc. A Q-switched Nd-YAG laser (7\,ns pulses
with a variable energy, $\lambda =1064\rm \,nm$, 20\,Hz) was used as a
light source and, after a reflection on the sample, the filtered second
harmonic signal at 532\,nm was detected by a photomultiplier and boxcar
integrator.

Low-temperature IR  reflectivity measurements in the frequency range
30--670\,\cm\, (or, equivalently, 1--20\,THz) were performed using a Bruker
IFS-113v Fourier-transform IR spectrometer equipped with a liquid-He-cooled Si
bolometer (1.6\,K) serving as a detector. Room-temperature mid-IR
spectra up to 5000\,\cm{} were obtained using a pyroelectric deuterated triglycine sulfate detector.
THz complex transmission from 3 to 100\,\cm{} was measured using a
custom-made time-domain spectrometer utilizing a Ti:sapphire femtosecond laser.
For the low-temperature IR reflectivity and THz complex transmission
spectroscopy, Oxford Instruments Optistat continuous-He-flow cryostats with polyethylene and mylar
windows, respectively, were used. THz spectra with magnetic field were measured
in an Oxford Instruments Spectromag cryostat with a superconducting magnet,
capable of applying an external magnetic field of up to 7\,T. The spectra were
recorded in the Voigt geometry with both polarizations, i.e., the electric vector of
the THz radiation  parallel or perpendicular to the external magnetic field.

For Raman studies, a Renishaw RM\,1000 Micro-Raman spectrometer equipped with a
CCD detector and Bragg filters was used. The experiments were performed in the
backscattering geometry within the 10--800\,\cm\ range using an Ar$^{+}$ ion laser with
the wavelength of 514.5\,nm and an Oxford Instruments Optistat optical
continuous-He-flow cryostat. Further, using a Quantum Design MPMS XL 7T and PPMS 9T instruments, we
carried out measurements of the magnetic susceptibility, magnetization and heat
capacity in the temperature interval of 2--400 K.

Magnetic resonance measurements were performed using a conventional electron
paramagnetic resonance (EPR) spectrometer operating at 9.2--9.8\,GHz and
within $T=10$--570\,K. An Oxford Instrument cryostat was used for
temperature measurements in the range from 10 to 300\,K. The  measurements at $T>300$\,K utilized a nitrogen-gas
variable temperature control system.

\section{Results and discussion}

\subsection{Structural properties}

The Rietveld analysis of our XRD data taken at various temperatures down to 5\,K
(see Figs.~2--4 in the Supplemental Material [\onlinecite{Suppl-2H-BaMnO3}]) confirmed
the structural phase transition at $\Tc=130$\,K. It manifests itself mainly by a
negative thermal expansion coefficient of the $\mathbf{c}$ lattice parameter
below $\Tc$ (see inset in Fig.~\ref{fig:XRD-RT}), which was observed recently also by
Stanislavchuk \textit{et al.}\cite{Stanislavchuk15} using neutron diffraction.
We would like to emphasize that a similar negative thermal
expansion of the lattice was observed below improper FE phase
transitions in YMnO$_3$ and HoMnO$_3$,
respectively,\cite{Gibbs11,Selbach12} which have the same space groups as 2H-BaMnO$_3$.

In our XRD patterns, below $\Tc$, no satellites expected for a tripled unit cell
were detected, due to their extreme weakness. Nevertheless, based on our results
of electron diffraction, we refined the low-temperature structure in the
$P6_{3}cm$ space group with the unit cell tripling in the hexagonal plane. In
the high-temperature phase, the structure refinement was possible in  polar
$P6_{3}mc$ and non-polar $P6_{3}/mmc$ space groups ($Z=2$) with almost the same
R factors within the error margin (see Figs.~4 in the Supplemental Material [\onlinecite{Suppl-2H-BaMnO3}]). Therefore, based on the XRD measurements
alone, we cannot decide whether the structure is polar or not.

\begin{figure}
            \centering
            \includegraphics[width=86mm]{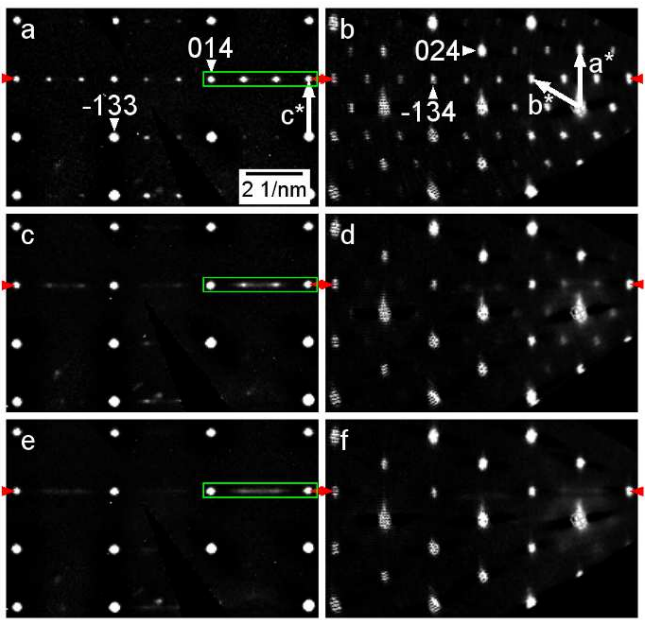}\\
	    \caption{Reciprocal space cuts through (a)
	      ($h+\frac{1}{2}$)(-2h)l and (b) $hk4$ planes
	     measured by electron
	    diffraction tomography at 99\,K. The cut planes are
	    perpendicular to each other, intersecting at the line marked by red
	    arrows. Parts (c)--(f): same cuts obtained at 130\,K and 170\,K. The green
	    boxes indicate the area used to calculate the intensity
	    profiles in
	    Fig.~\ref{fig:el-dif-intensity}.}
            \label{fig:EDT}
    \end{figure}

\begin{figure}
            \centering
            \includegraphics[width=80mm]{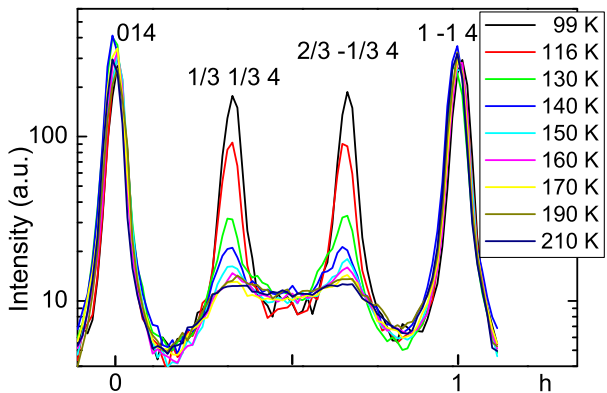}
	    \caption{Electron diffraction intensity profiles of the four
	      reflections along the $hk4$ line
	      ($k=1-2h$) at different temperatures. The
	  inner two peaks reveal the superstructure.}
            \label{fig:el-dif-intensity}
    \end{figure}

Electron diffraction was measured down to 99\,K. At each temperature the
crystal structure was refined using the full-matrix least-square refinement
and using the dynamical diffraction theory to calculate model intensities. The
   diffraction patterns from 140 to 240\,K correspond to 2H-BaMnO$_3$ without any presence of other polytypes
  (see Fig.~\ref{fig:EDT}). Below \Tc, the crystal undergoes a reversible
  phase transformation
  to the $P6_{3}cm$ space group with a tripled unit cell (see
  the satellite reflections in Figs.~\ref{fig:EDT} and
  \ref{fig:el-dif-intensity}), in agreement with
  Ref.~\onlinecite{Stanislavchuk15}. No
  other remarkable features were detected in the diffraction patterns.
  The intensities of the
  superlattice reflections in the \textit{hk0}
  plane (not shown) are negligible, implying that the polarization-inducing atomic shifts of Ba cations against the MnO$_6$
  octahedra at the phase
  transition  occur predominantly along the \textit{c} axis.

The spontaneous polarization of the low-temperature phase is related to the
shifts of the cations out of the high-symmetry positions. Assuming that
Mn atoms occupy exactly the centers of the MnO$_6$ octahedra,
the polarization is proportional to the mean difference of the shifts of Ba and
Mn atoms from their high-symmetry positions. This offset equals
0.0079\,$\mathbf{c}$ at 99\,K, 0.0080\,$\mathbf{c}$ at 116\,K, and
0.0093\,$\mathbf{c}$ at 130\,K. At the same time, the relative shift of the
Mn(1)O$_3$ and Mn(2)O$_3$ chains along \textit{c} decreases from
0.0284\,$\mathbf{c}$ at 99\,K to 0.0214\,$\mathbf{c}$ at 116\,K and
0.0157\,$\mathbf{c}$ at 130\,K. Thus, whereas the  shift of Mn atoms
decreases towards the phase transition, the estimated polarization remains
approximately constant and does not vanish at \Tc. This indicates that the
high-temperature phase could be also polar.

On heating, the intensities of the superlattice reflections
gradually decrease (see Fig.~\ref{fig:el-dif-intensity}) and they level off above 170\,K. Simultaneously,
additional diffuse scattering appears above \Tc, testifying to
short-range (dynamic) nanoregions with a polar
  $P6_{3}cm$ symmetry and a tripled unit cell, persisting at least up to
  170\,K. The  diffuse scattering can be explained by a mutual
  shift of the nearest columns of MnO$_6$ octahedra whereas
  the next-nearest ones are shifted reversely. Above \Tc, if the
    weak superlattice reflections are neglected, the structure can be refined
    again in both
  $P6_{3}/mmc$ and $P6_{3}mc$ space groups with almost the same
  figures of merit and almost identical structure models. In absence
  of inversion twins, this would indicate that the structure is
  centrosymmetric, as there is no evidence for symmetry lowering.
  However, we cannot exclude the presence of nanodomains twinned by inversion,
  so we cannot distinguish whether the high-temperature phase is polar or
  unpolar.

SHG is a powerful method for determining
polar order,
because it senses the second-order susceptibility which is non-zero only
in non-centrosymmetric crystal
structures. The results were, however, not quite conclusive. Due to a high
absorption and a low
damage threshold we could
employ only rather weak pulses, which reduced the detection
sensitivity. At room temperature,
we were able to detect a
signal somewhat stronger than the noise level. However, any surface
of the sample is non-centrosymmetric, so it can also contribute to the
detected SHG signal. In this case it was difficult to separate both possible
contributions.  On cooling, the signal remained almost constant, only
around $\Tc\approx 130\,\rm K$ a small increase was observed;
this was, unfortunately, not quite reproducible. Altogether, the
quality of the SHG data was insufficient to determine reliably whether the
high-temperature phase is non-centrosymmetric. For that reason,
  below we discuss our results considering both options,
  i.e., that the phase above \Tc{} may be polar or non-polar.

\subsection{Dielectric properties}
Assuming that both phases are polar (space groups $P6_{3}mc$ and
$P6_{3}cm$ for the high- and the low-temperature one, respectively), one may ask
whether these phases are FE, i.e., if they have a switchable
polarization. To investigate this issue, we have performed detailed
dielectric and lattice-dynamics studies.

\begin{figure}
            \centering
            \includegraphics[width=80mm]{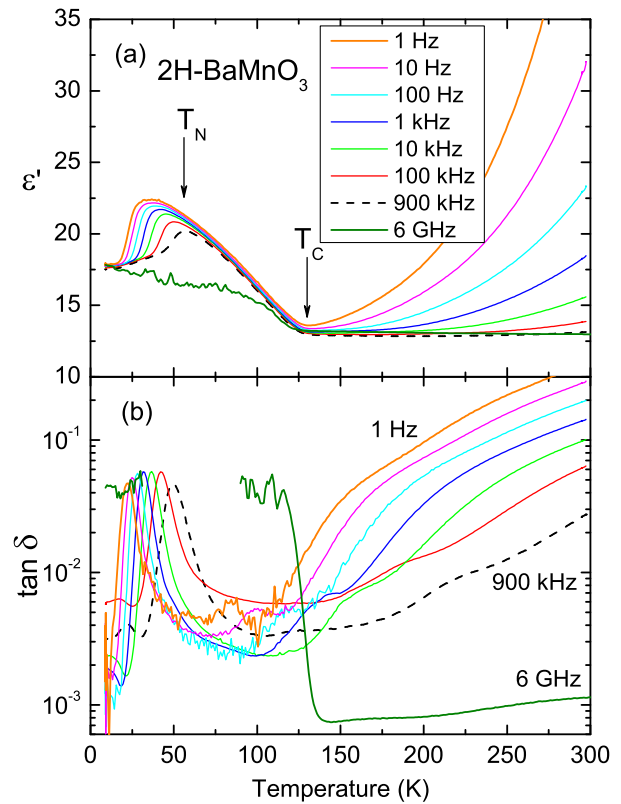}\\
	    \caption{Temperature dependence of (a) dielectric permittivity and
	    (b) dielectric loss of 2H-BaMnO$_3$ ceramics measured at various
    frequencies. The values of $\tan\delta$ at 6\,GHz in the temperature interval between 30
    and 90\,K are missing due to high losses.}
            \label{fig:BaMnO3_eps-T}
    \end{figure}

The temperature dependence of the permittivity $\varepsilon^{\prime}(T)$ shows a rather unusual behavior:
upon cooling, $\varepsilon^{\prime}$ between 0.9\,MHz and 6\,GHz
is temperature independent down to \Tc, and below it increases---see
Fig.~\ref{fig:BaMnO3_eps-T}. The absence of a peak  at \Tc{} and the  gradual increase in $\varepsilon^{\prime}$ below \Tc\, are typical
  of nonferroic
  second order phase transitions (see Fig.\,1 in Ref.~\onlinecite{Toledano82}) which supports the
  conjecture of
  a transition between two polar phases $P6_{3}mc$ $(Z=2)$ and $P6_{3}cm$ $(Z=6)$.
  A hypothetical phase transition from a nonpolar $P6_{3}/mmc$
  $(Z=2)$ to a $P6_{3}cm$
  $(Z=6)$ phase should be improper FE\cite{Varignon13} and, in this case,
  a jump or a small peak in $\varepsilon^{\prime}$ would be expected at \Tc.
  \cite{Levanyuk74,Dvorak74}
Below $\approx 60$\,K, $\varepsilon^{\prime}$ decreases, but this decrease is
strongly frequency dependent, reminding of a relaxor FE behavior.
This dielectric dispersion is apparently caused by slowing down of
domain-wall vibrations, as the relaxation times $\tau$ obtained from
the maxima of dielectric losses $\varepsilon^{\prime\prime}(f,T)$ [see Fig.~\ref{fig:BaMnO3_eps-T}(b)]
follow the Arrhenius law $\tau=\tau_{0} \exp\frac{E_{\rm a}}{k_{\rm B}T}$ with the
following parameters: $E_{\rm a}=0.04803\,\mbox{eV}$ $(557.4\,{\rm K})$ and $\tau_{0}=3.15 \times
10^{-12}$\,s (see Fig.~5 in the
Supplemental Material [\onlinecite{Suppl-2H-BaMnO3}]). A qualitatively similar effect was observed also in other
ferroelectrics.\cite{Huang97} The contribution of domain-wall vibration to
$\varepsilon^{\prime}$ was strongly reduced when it was measured in
a bias electric field of 11.3\,kV/cm (see Fig.~6 in the
Supplemental Material [\onlinecite{Suppl-2H-BaMnO3}]).

The dielectric dispersion observed above \Tc{} at low frequencies is
caused by the Maxwell-Wagner polarization, due to different values of
conductivity of grains and grain boundaries in the
ceramics.\cite{Lunkenheimer02}

\begin{figure*}
  \parbox[b]{\columnwidth}{\includegraphics[width=70mm]{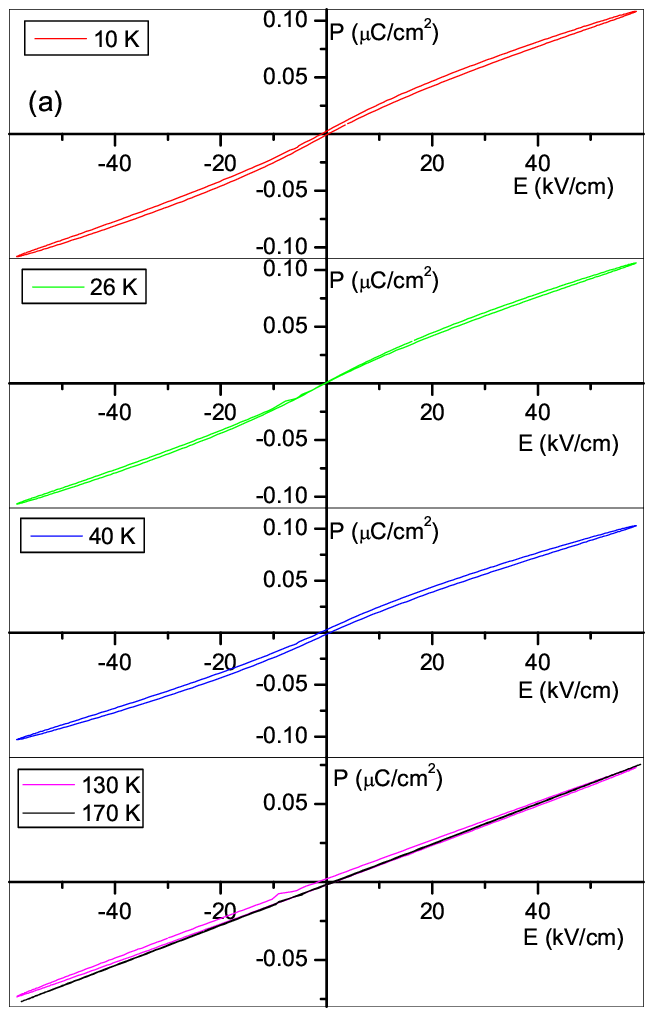}}
  \parbox[b]{\columnwidth}{\includegraphics[width=77mm]{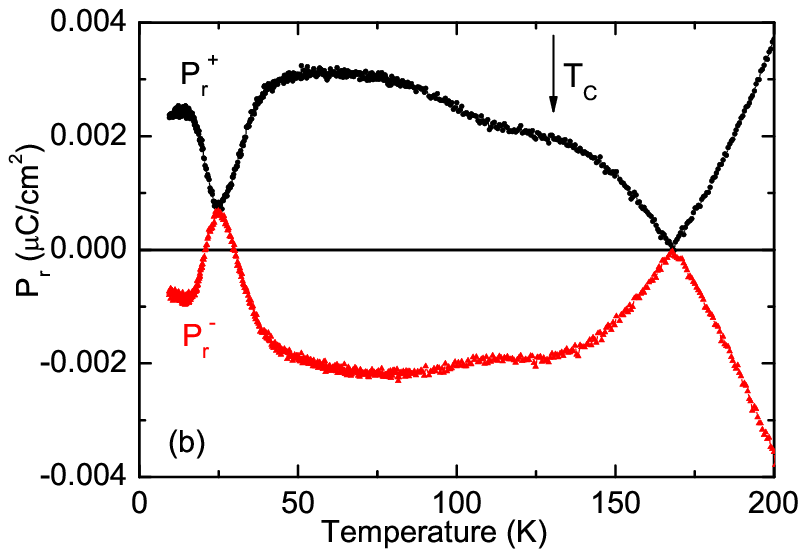}\\[2ex]
            \includegraphics[width=75mm]{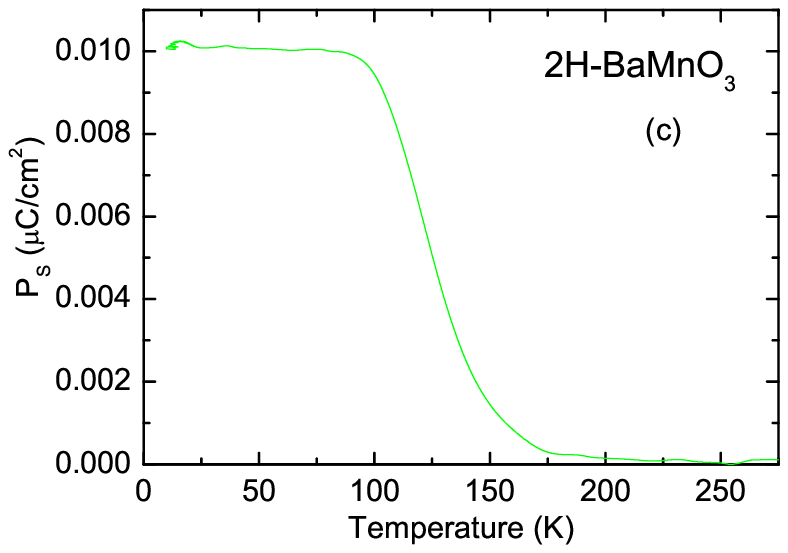}}
	    \caption{(a) FE hysteresis loops taken at selected temperatures.
	      Note the different scale for $P$ at $T=130\,\rm K$ and 170 K; The
	      latter one is paraelectric, i.e., not opened.
		    Temperature dependence of (b) remnant $P_{\rm
		    r}(T)$ and (c) spontaneous $P_{\rm s}(T)$ polarization
		  calculated from pyrocurrent after subtraction of the assumed thermo-stimulated
	      current due to defect migration. Here we would like stress, the polarization is non-zero in the high-temperature phase, here plotted $P_{\rm s}(T)$ is just enhanced polarization due to phase transition to low-temperature phase.
}
            \label{fig:polarization}
    \end{figure*}

Polarization hysteresis-loop measurements were performed in a
sinusoidal AC field. At higher temperatures, they revealed lossy elliptic
loops due to conductivity, down to 170\,K when the loops become a straight
line typical of paraelectrics. Below 170\,K the loops open again,
but although they have sharp ends, they do not saturate. Below \Tc, the loops have a slim S-shape [see
  Fig.~\ref{fig:polarization}(a)] typical of relaxors or weak ferroelectrics.
  The remnant polarization $P_{\rm r}\equiv$$P(E=0)$ exhibits an
  unusual temperature behavior [see Fig.~\ref{fig:polarization}(b)].
  At higher temperatures, it shows high values due to extrinsic
  conductivity. Upon cooling,  $P_{\rm r}$ drops and it becomes zero at 170\,K
  when the paraelectric loop is seen. On further cooling,  $P_{\rm r}$ increases, saturates near \Tc{} and increases
    again. Below $\approx\,60$\,K it starts to decrease
and, finally, it becomes pinched (i.e., $P^+_{\rm r}=P^-_{\rm r}$; here $P^+_{\rm
r}$ and $P^-_{\rm r}$ mark the remnant polarization for electric field going to zero from positive and negative sites, respectively) at
$T^{*}=26$\,K in zero electric field and opens only for $E\neq0$, which reminds
of antiferroelectrics. On subsequent cooling the hysteresis
loop opens again.

What is the origin of this effect? Constrained hysteresis
loops were reported already in many doped ferroelectrics like hard Pb(Zr,Ti)O$_3$
\cite{Carl78,Jin14} and they were explained by an internal bias field $E_i(t)$ due to defects.
According to Ref.~\onlinecite{Carl78}, this field increases with doping, but the
hysteresis-loop distortion disappears after a repeated cycling. This so called
``hysteresis relaxation'' obeys a time law in the form $E_i(t)\propto
\exp(-t/\tau)$ and it is found to be a both field- and thermally
activated process.\cite{Carl78} However, the behavior observed in
2H-BaMnO$_3$ is different: the pinched hysteresis loops at 26\,K remain
unchanged after many cycles (measured at 50\,Hz for 10 min.). Moreover, they
open again at lower temperatures, which would be excluded by the
above-mentioned mechanism which assumes a thermally stimulated process.

Levanyuk and Sannikov investigated theoretically improper ferroelectrics using
the phenomenological Landau theory\cite{Levanyuk74} and, based on the
temperature dependencies of the coefficients in the Landau expansion, they
predicted that the hysteresis loops may get pinched and take up shapes similar
to antiferroelectric ones. Nevertheless, they did not predict
a reentrant phase transition to a FE state with a single open
hysteresis loop. Moreover, for 26\,K, Fig.~\ref{fig:polarization} shows a
double hysteresis loop which is not one typical of
antiferroelectrics with
some critical electric fields $E_{\rm c1}$ and $E_{\rm c2}$; our loop is open at
any $E\neq0$. Interestingly, the loop is pinched at $E=0$, but we observed
that $P^+_{\rm r}=P^-_{\rm r}\neq0$ (Fig~\ref{fig:polarization}b).
We are not aware of any published
report of this effect, which can be explained probably
by the fact that few improper ferroelectrics were studied up to
now and also by the low values of $P_{\rm s}$ in these materials, which
make it difficult to measure
the FE loops directly. Most probably, the pinching is caused by some defects,
but we cannot exclude some intrinsic origin.

At all temperatures, the loops are very slim and the polarization
is very small, which can rise doubts about the ferroelectricity in 2H-BaMnO$_3$.
For that reason we have measured the pyroelectric current [zero-field
heating after a cooling at 40\,kV/cm, (see Fig.~7(a) in the
Supplemental Material [\onlinecite{Suppl-2H-BaMnO3}])] from which we calculated the
   spontaneous
polarization $P_{\rm s}(T)$ [see Fig.~\ref{fig:polarization}(c)]. One can
see the onset of $P_{\rm s}$ below 175\,K, its increase at $T_{C}$ and
saturation below 100\,K. The non-zero values of $P_{\rm s}$ and $P_{\rm
r}$ above \Tc{} can be explained by the persistence of clusters of
the low-temperature phase above \Tc{}, which is confirmed also by the electron diffraction.

Nevertheless, we should note that the $P_{\rm s}(T)$ was calculated from the
pyrocurrent by subtracting both high- and low-temperature increase (see Fig.~7 in the
Supplemental Material [\onlinecite{Suppl-2H-BaMnO3}]), which we attribute to thermally stimulated current of migrating
defects (mainly oxygen vacancies and related Mn$^{3+}$). If we subtract only high-temperature contribution, the calculated polarization
would be unrealistically one order of magnitude higher. In such a case, $P_{\rm s}$ would linearly
increase on cooling below 180\,K (as much as 50\,K above \Tc, see Fig.~7(b) in the
Supplemental Material [\onlinecite{Suppl-2H-BaMnO3}]),
exhibiting no anomaly near \Tc. Such a linear dependence of $P_{\rm s}(T)$ can
be expected below an improper FE phase transition,\cite{Levanyuk74,Dvorak74} but
in that case the unit cell should triple already at 180\,K. One can argue that
\Tc{} could be shifted by the external electric field of 40\,kV/cm{} used before
the pyrocurrent measurements. However, we also investigated
$\varepsilon^{\prime}(T)$ with an external electric field of 11.3\,kV/cm and we
observed no shift in \Tc. For that reason we believe that the current detected
below 100\,K is a thermally stimulated one which, for a proper polarization
calculation, should be subtracted, and that the phase transition at $\Tc=130\,$K
is nonferroic between two polar phases.

\subsection{Phonon spectra}

Based on the diffraction experiments alone, we cannot unambiguously
  distinguish between nonpolar $P6_{3}/mmc$ and polar $P6_{3}mc$ space groups in
  the high-temperature phase. Lattice-vibrations spectra can yield useful
  information, because the phonon selection rules are different in
  different space groups.  We have performed a factor-group analysis of phonons, where we used site symmetries of all atoms known from structural refinements\cite{Cussen00} and group tables in
Ref.~\onlinecite{Rousseau81}. If, above \Tc, 2H-BaMnO$_3$ crystallizes in
the $P6_{3}/mmc$ $(D_{6h}^{4})$ structure with $Z=2$, the zone-center phonons have the following symmetries:
\begin{eqnarray}
\Gamma_{D_{6h}^{4}}&=3A_{2u}(z)+4E_{1u}(x,y)+2E_{2u}+2B_{2u}+\nonumber
\\ &+A_{1g}(x^2+y^2,z^2)+A_{2g}+2B_{1g}+B_{1u}+\nonumber
\\ &+E_{1g}(xz,yz)+3E_{2g}(x^2-y^2,xy). \end{eqnarray}
Here $x$, $y$, and
$z$ in parentheses mark the electric polarizations of the IR radiation for which the
phonons are IR active, whereas the rest of the symbols in parentheses
are components of the Raman tensor. Phonons without parentheses are
silent. Except for two acoustic phonons ($A_{2u}+E_{1}$),
five IR-active and another five Raman-active phonons can be expected in the spectra.

If the high-temperature phase has a polar $P6_{3}mc$ $(C_{6v}^{4})$
structure with
$Z=2$, the phonons from the $\Gamma$ point of
the BZ have the following symmetries :
\begin{eqnarray}
\Gamma_{C_{6v}^{4}}&=4A_{1}(z,x^2+y^2,z^2)+5E_{1}(x,y,xz,yz)+A_{2}+\nonumber
\\ &+B_{1}+4B_{2}+5E_{2}(x^2-y^2,xy)\label{eq:fga_polarhi}\end{eqnarray}
After subtracting two acoustic phonons, 7 IR- and Raman-active and
five only
Raman-active phonons are predicted in the spectra.

A phase transition connected with a tripling of the unit cell (and folding of
the BZ)
should exhibit a noticeable change of the selection rules for activation of phonons in
IR and Raman spectra.
The factor-group analysis of phonons in the low-temperature $P6_{3}cm$
($C_{6v}^{3}$) structure with $Z=6$ yields:
\begin{eqnarray}
\Gamma_{C_{6v}^{3}}&=9A_{1}(z,x^2+y^2,z^2)+15E_{1}(x,y,xz,yz)+6A_{2}+\nonumber
\\ &+9B_{1}+6B_{2}+15E_{2}(x^2-y^2,xy). \end{eqnarray}
In this case,
after subtracting two acoustic phonons, 22 modes ($8A_{1}+14E_{1}$)
should be IR- and Raman-active and additional 15 $E_{2}$-symmetry modes Raman-active only.

To compare the predictions with experiments, we measured Raman
scattering down to 5\,K (see Fig.~8 in the
Supplemental Material [\onlinecite{Suppl-2H-BaMnO3}]) and obtained spectra similar to those reported by Stanislavchuk
\textit{et al.}\cite{Stanislavchuk15} Namely, we observed 6 modes at 300\,K and only 8
sharp modes at 5\,K, i.e., one more than the number allowed in
the $P6_{3}/mmc$ phase and much fewer than  allowed in the
FE phase (the mode frequencies are listed in Table 1). The only
  newly observed mode, activating below \Tc{} near 15\cm{} and
  hardening to 26\cm{} on cooling, could not be seen in
  Ref.~\onlinecite{Stanislavchuk15} where the Raman spectra were measured only above 100\cm.

\begin{figure}
            \centering
            \includegraphics[width=85mm]{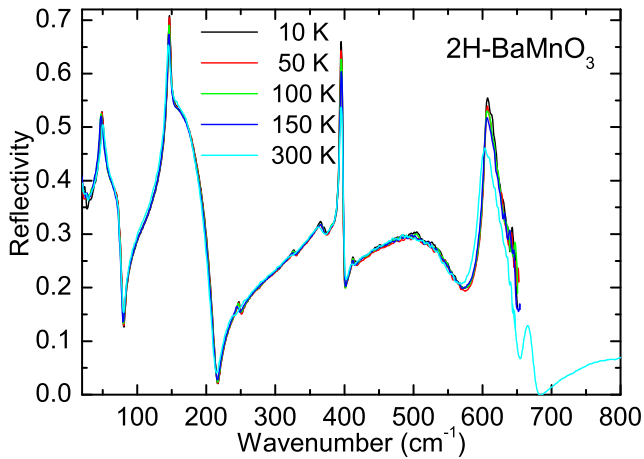}
            \caption{IR reflectivity spectra of 2H-BaMnO$_3$ ceramics taken at
	    various temperatures. The upper frequency limit of the low-temperature
    spectra is determined by the transparency of bolometer and cryostat windows.}
            \label{fig:IRspectra}
\end{figure}
Fig.~\ref{fig:IRspectra} shows the IR reflectivity spectra of
2H-BaMnO$_3$ ceramics which, surprisingly, exhibit no dramatic changes with
temperature. Some modes sharpen slightly on cooling due to their decreasing
damping, but the number of IR-active phonons does not increase below \Tc, in
contrast to expectations based on the selection rules discussed above.
This must be related to the fact that the FE distortion is very small and
the newly allowed modes in the FE phase are very weak and therefore
not detectable in the IR and Raman spectra.

\begin{figure}
            \centering
            \includegraphics[width=80mm]{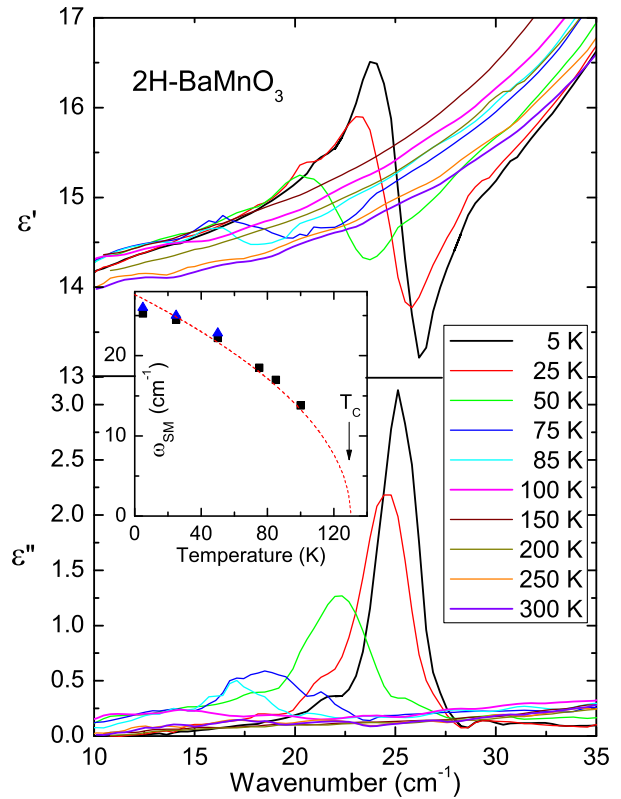}
            \caption{
	      THz complex dielectric spectra of 2H-BaMnO$_3$ ceramics,
		    showing activation of a new phonon below \Tc{} = 130\,K. Inset shows the
	    temperature dependence of the soft-mode frequency obtained from the THz (squares) and Raman (triangles) spectra. Dashed line is the result of the fit using Cochran law.}
            \label{fig:THz}
\end{figure}

We measured the IR-active phonons also using the more sensitive time-domain THz
transmission spectroscopy. As the sample was opaque between 50 and 80\,\cm,
where a strong reflection band is seen in Fig.~\ref{fig:IRspectra}, we present
the THz spectra only below 35\,\cm{} (see Fig.~\ref{fig:THz}); they reveal the
same low-frequency mode as the Raman spectra. This weak mode
appears at 100\,K near 13\,\cm{} and it hardens on cooling, reaching the
frequency of 26\,\cm{} at 5\,K. Its temperature dependence (see inset of
Fig.~\ref{fig:THz}) follows the Cochran
law, $\omega_{\rm SM}=A\sqrt{\Tc-T}$ with $\Tc=130$\,K, implying that this is the soft mode driving the structural
phase transition with unit cell tripling. Above \Tc, this phonon has a
wavevector $q = (\frac{1}{3},\frac{1}{3},0)$ at the BZ boundary, so it
cannot be observed in the IR or Raman spectra, which  detect only phonons with $q = 0$. Below \Tc{} the BZ folds, the phonon wavevector transforms to the BZ center and  the soft mode activates in
the THz and Raman spectra.

The complex dielectric spectra shown in Fig.~\ref{fig:THz} and the IR
reflectivity spectra in Fig.~\ref{fig:IRspectra} were fitted using a model
consisting of a sum of damped Lorentzian oscillators.\cite{Buixaderas04} The
dielectric strength $\Delta\varepsilon_{SM}$ of the soft mode was found to be
0.2 at \textit{T}=5\,K and this value significantly decreased with increasing
temperature. Thus, the contribution of the soft mode to the static permittivity
is one order of magnitude lower than that of domain wall vibrations (5--10), but
this fact is well known also from other ferroelectrics like PZT\cite{Porokhonskyy09} and BaTiO$_3$\cite{Choi10}.

It is possible to show that the soft mode cannot be due to an AFM resonance. In
fact, we measured THz spectra with an external magnetic field of up to
7\,T, which showed that the mode does not change with the field. This result
questions a magnetic origin of the soft mode and supports its phonon origin.
The absence of an AFM resonance in THz spectra is in agreement with the
earlier statement by Christensen \textit{et al.}\cite{Christensen72} about the same size of the
structural and magnetic cells ($Z=6$). The
stability of the phonon parameters and of the THz permittivity with respect to
external magnetic field also confirms a negligible magnetodielectric coupling in 2H-BaMnO$_3$.

We are coming back to the question whether the phase transition at \Tc{}
is an improper-FE one, from a non-polar
$P6_{3}/mmc$  to a polar
$P6_{3}cm$  structure, or a nonferroic one between two
polar space groups $P6_{3}mc$  and $P6_{3}cm$. Since
the diffraction experiments could not clearly distinguish between
these two cases, older works \cite{Christensen72,Cussen00,Varignon13}
claimed a non-polar symmetry at room temperature. Some newer
papers\cite{Roy98,Stanislavchuk15}, combining structural and lattice vibration
studies, came to the conclusion that the high-temperature phase is polar,
because two of the six observed phonons are both IR- and Raman-active.

\begin{table*}
\caption{Comparison of mode frequencies (in \cm) observed in our IR and Raman spectra at
  room temperature and 10\,K and those published in Ref.~\onlinecite{Stanislavchuk15}, along with their
    symmetry assignments in polar space groups. Symmetry of heavily
    damped modes coming from crystal field or defects is not assigned. The modes in the
high-temperature phase which cannot be accounted for by the factor-group
analysis in Eq.~(\ref{eq:fga_polarhi}), are marked by stars.}

  \medskip
\begin{tabular}{lrcccccccc}
\hline
\hline
  & & \multicolumn{4}{c}{300\,K} & \multicolumn{4}{c}{10\,K} \\
\cline{3-6} \cline{7-10}
&&\small ours &\small Ref.~\onlinecite{Stanislavchuk15}&\small  ours &\small Ref.~\onlinecite{Stanislavchuk15}  &\small ours & \small Ref.~\onlinecite{Stanislavchuk15} &\small  ours &\small Ref.~\onlinecite{Stanislavchuk15}  \\
symmetry & note \hspace*{5mm} & $\omega_{\rm IR}$ & $\omega_{\rm IR}$ & $\omega_{\rm Raman}$ & $\omega_{\rm Raman}$ & $\omega_{\rm IR}$ & $\omega_{\rm IR}$ & $\omega_{\rm Raman}$  & $\omega_{\rm Raman}$   \\
\hline
\hline
 $E_{1}$ & & -- & -- & -- & --& 26 & -- & 26 & -- \\
 $A_{1}$ & & 50 & 82 & -- & --& 50 & 82 & -- & -- \\
 ? & heavily damped & 65 & -- & -- & --& 65 & -- & -- & -- \\
 $E_{2}$ & & -- & -- & 119.5 & 119& -- & -- & 120.3 & 120 \\
 $E_{1}$  & & 143 & 145 & -- & -- & 143 & 146.3 & -- & --\\
 ? & heavily damped & 156$^{*}$ & -- & -- & -- & 156 & -- & -- & -- \\
 ? & sharp and weak & 246$^{*}$ & -- & -- & -- & 246 & -- & -- & -- \\
 ? & heavily damped& 294$^{*}$ & -- & -- & -- & 294 & -- & -- & -- \\
 ? & sharp and weak & 327$^{*}$ & -- & -- & -- & 327 & -- & -- & -- \\
 ? & heavily damped & 331$^{*}$ & -- & -- & -- & 331 & -- & -- & -- \\
 $E_{2}$ & & -- & -- & -- & 344 & -- & -- & -- & 344\\
 $A_{1}$ & & 362 & -- & -- & -- & 362 & -- & 369 & 368\\
 ? & heavily damped & 378$^{*}$ & -- & -- & -- & 378 & -- & -- & -- \\
 $E_{1}$  & & 394 & 394.5 & -- & -- & 394 & 394 & -- & --\\
 $E_{1}$ & & 410 & -- & 414.5 & 413 & 412 & -- & 417 & 417\\
 ? & heavily damped & 460$^{*}$ & -- & -- & -- & 460 & -- & -- & -- \\
 $A_{1}$ & & -- & -- & 490 & 487 & -- & -- & 490 & 492\\
 ? & heavily damped & 499$^{*}$ & -- & -- & -- & 499 & -- & -- & -- \\
 $E_{2}$ & & -- & -- & 527 & 526 & -- & -- & 527 & 526 \\
 $E_{1}$  & & -- & 599 & -- & -- & 593 & 605 & -- & --\\
 $A_{1}$ & & 637 & 639 & 639  & 639 & 637 & 645 & 644  & 639\\
 $E_{1}$  & & 663 & 653 & 660 & 660 & -- & 653 & 661 & 660\\
\hline
\hline
  \end{tabular}
\end{table*}

In Table 1 we present mode frequencies observed in our IR and Raman spectra
compared with  those published previously\cite{Stanislavchuk15}.
At room temperature, we can see only three out of seven modes allowed in
the high-temperature polar phase, which are simultaneously IR and Raman active
(at 410, 637 and 663\,\cm). However, we detected altogether 16 IR-active modes,
which significantly exceeds the number  allowed in both possible
high-temperature phases. Seven heavily damped modes (marked in Table~1)
exhibit a constant damping on cooling, which is not a behavior
typical of phonons.  These broad bands can be crystal-field excitations of
Mn$^{4+}$,\cite{Nalecz16} or some defect-induced modes. The shape of the
  broad IR reflection bands reminds of spectra of crystalline
  glasses,\cite{Kratochvilova01} but our XRD did not reveal any traces of
secondary amorphous phases in our ceramics. Although these heavily damped modes
  were not observed in IR-ellipsometric spectra of a 2H-BaMnO$_3$ single
  crystal\cite{Stanislavchuk15},
  it is worth noting that these ellipsometric spectra were taken
  mostly in the hexagonal plane (i.e., sensitive to $E_{1u}$ or $E_{1}$
  symmetry spectra). Thus, the broad modes are probably active in
  the \textbf{E}$\parallel$\textbf{c}-polarized spectra. After
    subtracting these seven modes, nine of them remain, still more than allowed
    by the factor-group analyses in both possible high-temperature
    symmetries. The excess modes can be explained by the presence of
    dynamical nanoregions (as seen also in the electron
    diffraction above \Tc) with a tripled unit cell; it is known
    that IR spectroscopy is extremely sensitive to a locally broken
    symmetry. Similar effects are well known also from relaxor
    ferroelectrics.\cite{Hlinka06}

In summary, it is not easy to identify clearly the space group at $T>\Tc$ as
polar or nonpolar, based on IR and Raman spectra only. Also our structural,
SHG and dielectric studies yielded no unambiguous result---the
high-temperature phase can be either polar or nonpolar. In any case, the
structural phase transition at \Tc{} is non-equitranslational. We have
discovered the soft mode which drives the phase transition;
observations of this kind have been reported only rarely up to now. Previous lattice-dynamics studies of
improper ferroelectrics revealed either no soft
modes\cite{Petzelt73,Lockwood76} or merely an overdamped excitation resolved as
a central mode\cite{Petzelt84}. Even in hexagonal RMnO$_3$ manganites
($\rm R=Sc$, Y, Dy--Lu) which exhibit improper
FE phase transitions\cite{Lilienblum15,Fennie05} analogous to
2H-BaMnO$_3$, no soft modes were reported.\cite{Goian10a,Bouyanfif15}

Concerning non-ferroic phase transitions between two polar phases, very few materials with
such phase transitions are known (examples include
PbZr$_{x}$Ti$_{1-x}$O$_{3}$, (NH$_{4})_{2}$H$_{3}$IO$_{6}$ or
Na$_{1-x}$K$_{x}$NbO$_3$).\cite{Toledano82} They exhibit
  dielectric and phonon properties different from 2H-BaMnO$_3$---neither an underdamped soft mode nor an increase in $\varepsilon^{\prime}$
   were observed below \Tc. Only in one case, activation of new hard phonons due to
unit-cell multiplication was reported.\cite{Buixaderas16}  To our
knowledge, no report of a soft mode below a non-equitranslational
nonferroic phase transition between two polar phases has been
published yet.

\subsection{Landau theory}

The Landau theory of the improper FE phase transition in BaMnO$_3$
was published by Varignon \textit{et al.}\cite{Varignon13} who explained
the transition from a nonpolar $P6_{3}/mmc$ to polar $P6_{3}mc$ symmetry
by a coupling of a $K_3$-symmetry mode from the BZ boundary with a zone-center
mode of $\Gamma_2^{-}$ symmetry. In this case the free energy expansion in
  terms of their amplitudes $Q_{K_3}$ and $Q_{\Gamma_{2}^{-}}$ has the form\cite{Varignon13}
\begin{eqnarray}
F(Q_{K_3},Q_{\Gamma_{2}^{-}} )= &\alpha_{20}Q_{K_3}^{2}+\alpha_{02}Q_{\Gamma_{2}^{-}}^{2}+\beta_{40}Q_{K_3}^{4}+\nonumber
\\ &+\beta_{04}Q_{\Gamma_{2}^{-}}^{4}+\beta_{31}Q_{K_3}^{3}Q_{\Gamma_{2}^{-}}+\beta_{22}Q_{K_3}^{2}Q_{\Gamma_{2}^{-}}^{2}
\end{eqnarray}

Here $\beta_{31}$ and $\beta_{22}$ express coefficients of the cubic and biquadratic coupling, respectively.

Let us assume that the the non-equitranslational
phase transition in 2H-BaMnO$_3$ is nonferroic between two polar space groups
$P6_{3}mc$ and $P6_{3}cm$. In this case both phases exhibit a polarization along the
\textit{z} axis and the temperature dependence $P(T)$ in
Fig.~\ref{fig:polarization}(c) corresponds only to the increase in polarization
due to the structural phase transition.

For the wave vector $q = (\frac{1}{3},\frac{1}{3},0)$, the observed symmetry
change is induced by the two-dimensional irreducible representation $K_1$ of the $P6_{3}mc$ space group  which is associated with the following complex matrices:

\begin{eqnarray}
(C_{1},C_{3},C_{3}^{2},m_{x},m_{y},m_{xy})\rightarrow \left( \begin{array}{cc}
1 & 0 \\
0 & 1  \end{array} \right), \\
(C_{2},C_{6},C_{6}^{5},m_{1},m_{2},m_{3}|0,0,\frac{c}{2})\rightarrow \left( \begin{array}{cc}
0 & 1 \\
1 & 0  \end{array} \right), \\
\vec{t}_{1}=(a,0,0)\rightarrow \left( \begin{array}{cc}
\varepsilon^{2} & 0 \\
0 & \varepsilon  \end{array} \right), \\
\vec{t}_{2}=(0,a,0)\rightarrow \left( \begin{array}{cc}
\varepsilon & 0 \\
0 & \varepsilon^{2}  \end{array} \right), \\
\vec{t}_{3}=(0,0,c)\rightarrow \left( \begin{array}{cc}
1 & 0 \\
0 & 1  \end{array} \right)
\end{eqnarray}
with $\varepsilon=\exp(2\rm i\pi/3)$. Denoting the two-component order parameter
associated with $K_1$ as $\eta_{1}=\varrho \cos\theta$,
$\eta_{2}=\varrho \sin\theta$, the above matrices allow constructing the Landau free energy associated with the transition, which is
\begin{equation}
F= \frac{\alpha}{2}\varrho^{2}+\frac{\beta_{1}}{3}\varrho^{3}\cos3\theta+\frac{\beta_{2}}{4}\varrho^{4}+\frac{\gamma}{6}\varrho^{6}\cos^{2}3\theta .
\label{eq:F_ro_theta}
\end{equation}
Here a six-degree invariant has been included for a full description of all the phases of the phase diagram.

\begin{figure}
    \centering
    \includegraphics[width=0.8\columnwidth]{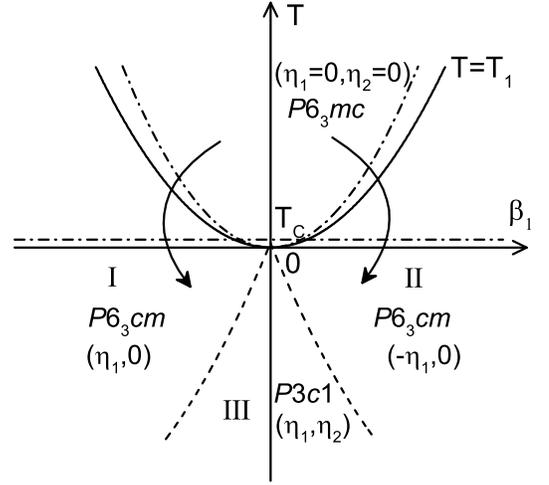}
    \caption{Phase diagram corresponding to the free energy $F$ given by
      Eq.~\ref{eq:F_ro_theta} in the $(T,\beta_1)$ plane.
Dashed and solid curves represent the second and first-order transition limits, respectively.
Dash-and-dot curves show the limits of stability. One of the two arrows indicates the
thermodynamic path followed at the reported transition in 2H-BaMnO$_3$.}
    \label{fig:phasdiag}
  \end{figure}

Minimizing $F$ with respect to $\theta$ and $\varrho$ one finds three possible
low-symmetry stable phases, denoted I, II and III (see Fig.~\ref{fig:phasdiag}).
Phases I and II, corresponding both to the equilibrium conditions
($\cos3\theta=\pm1, \varrho\neq0$), have the same symmetry $P6_{3}cm$ with a
tripling of the parent unit cell based on the basic translations
$2\vec{t}_{1}+\vec{t}_{2}$, $\vec{t}_{2}-\vec{t}_{1}$, $\vec{t}_{3}$. However,
they are anti-isostructural since one of them corresponds to ($\eta_{1}$,
$\eta_{2}=0$) and the other to (-$\eta_{1}$, $\eta_{2}=0$). These two phases are
stabilized with a fourth degree expansion of $F$. Phase III, which, for
$-1<\cos3\theta <1$, has a region of stability located between phases I and II,
has the $P3c1$ symmetry with the same tripling of the unit cell. This phase
requires a sixth-degree term in $F$ for its stabilization. Because of the cubic
term in $F$, the transition to the $P6_{3}cm$ phases I and II is necessarily of
the first order and takes place at a temperature $T_{1}>\Tc$. The transition to
the $P3c1$ phase III cannot occur directly from the parent phase, except at a
single point of the phase diagram.

The total polarization $P=P_{0}+\Delta P$ is left invariant by all the
symmetry operations of the $P6_{3}mc$ space group, which have the symmetry
of the $\Gamma_{1}$ BZ center representation. Therefore, under applied electric
field $E$, the dielectric free energy $F_{\rm D}$ has the general form:

\begin{equation}
F_{\rm D}= P\left (\frac{a}{2}\varrho^{2}+\frac{b}{3}\varrho^{3}\cos 3\theta \right)+\frac{\delta}{2}P^{2} \varrho^{2}+\frac{P^{2}}{2\chi_{zz}^{0}}-EP
\end{equation}
where $a$, $b$ and $\delta$ are coupling coefficients.
Taking into account the equilibrium condition $\cos 3\theta=1$  and
minimizing $F_{\rm D}$  with respect to $P$ yields the equation of state for  $P$ below $T_{1}$:

\begin{equation}
  P(\frac{1}{\chi_{zz}^{0}}+\delta \varrho^{2}) +
  \varrho^{2}(\frac{a}{2}+\frac{b}{3}\varrho) = E\,.
\label{eq1}
\end{equation}

Minimizing the total free energy $F+F_{\rm D}$, truncated at the fourth degree, with respect to $\varrho$ provides the other equation of  state:
\begin{equation}
\alpha+ \varrho(\beta_{1}+\beta_{2}\varrho)+ P(a+b\varrho+\delta P)=0.
\label{eq2}
\end{equation}

By extracting $\varrho$ as a function of $P$  from Eq.~(\ref{eq2}) and
replacing this expression into Eq.~(\ref{eq1}), one gets an equation
containing only $P$ and  $E$ which can yield the $E(P)$
dependence. Keeping all the terms in the equations requires numerical
simulations. An acceptable approximation providing an algebraic form for $E(P)$
can be obtained by neglecting the $\delta$, $b$  and $\beta_{2}$ terms in
Eq.~(\ref{eq2}), and the $\delta$ term in Eq.~(\ref{eq1}). This leads to:

\begin{eqnarray}
  E&=&\frac{P}{\chi_{zz}^{0}}+\frac{1}{\beta_{1}^{2}}
  \left [-\frac{a^3b}{3\beta_1}P^{3}+a^2\left(\frac{a}{2}-\frac{\alpha b}{\beta_{1}}\right)P^{2}+\right.\nonumber \\
 &+&\left. \alpha a\left(a-\frac{\alpha b}{\beta_{1}}\right)P +
 \alpha^2\left(\frac{a}{2}-\frac{\alpha b}{3\beta_{1}}\right)
 \right ]
\label{eq3}
\end{eqnarray}
from which one can deduce the dielectric susceptibility $\chi=\lim_{E\rightarrow0}\frac{\partial P}{\partial E}$:
\begin{equation}
  \chi=\frac{\chi_{zz}^0}{1+\frac{a\chi_{zz}^0}{\beta_1^2}\left[
    -\frac{a^2b}{\beta_1}P^2+a\left(a-2 \frac{\alpha b}{\beta_1}\right)P
    + \alpha \left(a-\frac{\alpha b}{\beta_1}\right)
  \right]}
\end{equation}
where $P$ takes its zero-field approximated value:
\begin{eqnarray}
  P=-\chi_{zz}^{0}(\frac{a}{2}\varrho^{2}+\frac{b}{3}\varrho^{3})
\label{eq5}
\end{eqnarray}
and $\varrho$ can be approximated (not too close to $T_{1}$) as $\varrho\approx(T_{1}-T)^{\frac{1}{2}}$.

Eq.~(\ref{eq5}), which constitutes a first approximation of $P(T)$, shows that depending on the sign and magnitude of the coefficients $a$ and $b$ and assuming a more accurate form for $\varrho(T)$, it may be possible to reproduce the observed behavior reported in Figs.~\ref{fig:BaMnO3_eps-T} and~\ref{fig:polarization}. Along the same line Eq.~(\ref{eq3}) also shows that a complicated $E(P)$ hysteresis behavior may occur.

\subsection{Magnetic properties} The earlier published reports about the AFM
phase transition in 2H-BaMnO$_3$ provided contradictory information. On the one
hand, a magnetic susceptibility $\chi(T)$ monotonically increasing on cooling
without any anomaly was observed by Chamberland $\textit{et
al.}$\cite{Chamberland70} and Christensen $\textit{et al.}$\cite{Christensen72}
This led them to the conclusion that \Tn{} was lower than 3\,K. On the other
hand, below $\Tn=59\pm2$\,K, Cussen and Battle\cite{Cussen00} observed, by means
of neutron diffraction, a long-range AFM order. Simultaneously, they found that
the magnetic susceptibility $\chi(T)$ does not follow the Curie-Weiss law,
below ca.\ 200\,K it exhibits different values in field- and zero-field regimes
and shows no maximum at \Tn.  They proposed to explain this behavior by
suggesting a short-range one-dimensionally ordered AFM phase at least up to
200\,K.
\begin{figure}
  \centering \includegraphics[width=80mm]{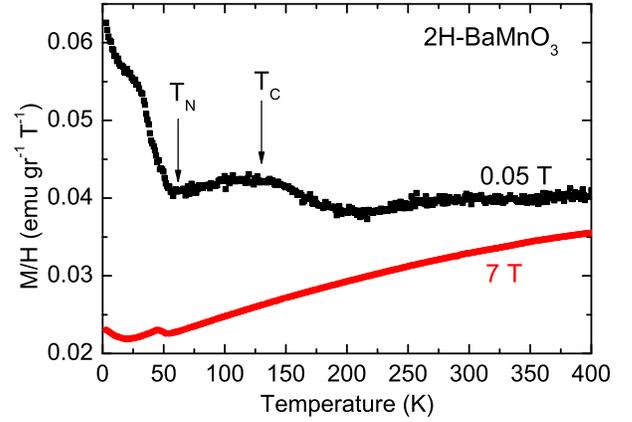}
  \caption{Temperature dependence of magnetization divided by magnetic field,
  measured with an external magnetic field of 0.05 and 7\,T. Note that neither
curve follows the Curie-Weiss law.}
\label{fig:magnetization}
\end{figure}

To shed more light on this issue, we performed low-temperature measurements of
magnetic properties of our samples.  The temperature dependence of $\chi$
measured at 0.05\,T revealed two minima near 60 and 220\,K (see
Fig.~\ref{fig:magnetization}). The first temperature corresponds well to \Tn{}
from Ref.~\onlinecite{Cussen00}; the second anomaly will be explained below. In
contrast, $\chi(T)$ measured at 7\,T shows a continuous decrease on cooling down
to \Tn. Neither the low- nor the high-field $\chi(T)$  follow the Curie-Weiss
law. Interestingly, our measurements of magnetization $M(H)$ up to $\mu_0 H =
7\,T$ did not reveal any saturation of magnetization even at 2\,K
(see Fig.~9 in the Supplemental Material [\onlinecite{Suppl-2H-BaMnO3}]).

\begin{figure}
            \centering
            \includegraphics[width=80mm]{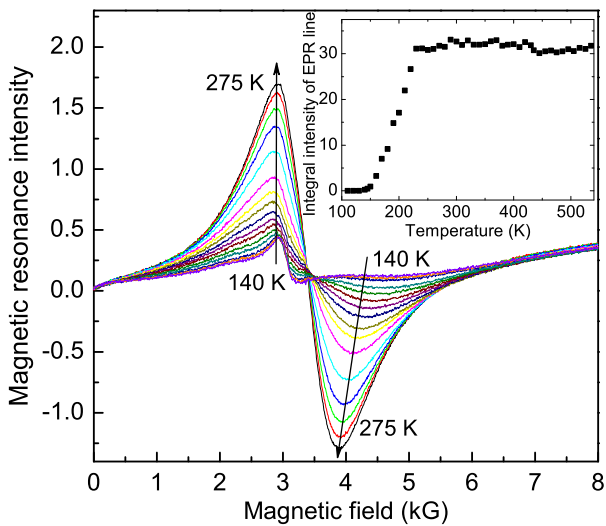}
            \caption{EPR
		    spectra of 2H-BaMnO$_3$
	    measured at 9.4\,GHz as a function of temperature.
	    Inset: temperature dependence of the integrated intensity of the EPR line.}
            \label{fig:EPR}
    \end{figure}

In order to identify the origin of such an unusual behavior, we performed
additional EPR measurements from 16 to 550\,K. The  EPR spectra measured in the
temperature region 140--275\,K are shown in Fig.~\ref{fig:EPR}. The only
resonance line can be attributed to the Mn$^{4+}$ ($S=\frac{3}{2}$) ions of
BaMnO$_3$. It has a Lorentzian shape with a peak-to-peak linewidth of 1100\,G at
275\,K which increases on cooling (see Fig.~10 in the Supplemental Material
[\onlinecite{Suppl-2H-BaMnO3}]). Note that a Lorentzian-shaped line with a similar
linewidth was observed in the paramagnetic phase of SrMnO$_3$.\cite{Alvarez08}
This suggests that the Mn$^{4+}$ EPR spectrum of BaMnO$_3$ is narrowed by the
exchange interaction, as expected for paramagnetic materials.\cite{Anderson53}
However, upon cooling below 230\,K, its intensity sharply decreases and the
spectral line almost disappears at $T\approx 150$\,K (see inset of
Fig.~\ref{fig:EPR}). Below this temperature only a weak residual spectrum
remains, slightly changing upon further cooling  and splitting below 80\,K (see
Fig.~11 in the Supplemental Material [\onlinecite{Suppl-2H-BaMnO3}]). This feature
is obviously linked to paramagnetic impurities at grain boundaries; note that it
amounts to about 0.5\% of the total Mn$^{4+}$ paramagnetic signal only, while
the main spectral line must originate from grains.

Further, we calculated the integral intensities via a double integration of the
Mn$^{4+}$ spectral line (see its temperature dependence in the inset of Fig.~\ref{fig:EPR}). As this intensity in the paramagnetic phase should be proportional to the magnetic susceptibility, one can see that the EPR line
intensity does not follow the Curie-Weiss law which would impose an increase on
cooling towards \Tn. Our spectral intensity is almost constant between 550 and
220\,K and it drastically decreases below $\approx$ 220\,K. This temperature dependence qualitatively corresponds to $\chi(T)$, where also no Curie-Weiss behavior was observed.

The decrease in the EPR intensity below 220\,K indicates a
gradual decrease in concentration of paramagnetic (magnetically uncoupled) Mn$^{4+}$
ions which develop a mutual AFM correlation
  at the EPR time scale ($10^{-9}$--$10^{-10}$\,s). Finally, as the
   absorption line completely disappears near 140\,K, all Mn$^{4+}$
spins become AFM-coupled with a zero total spin so that no paramagnetic absorption
is possible. This observation supports the hypothesis of Cussen and
Battle\cite{Cussen00} about the existence of a short-range one-dimensionally ordered
AFM phase below $\approx200\,$K. We can thus conclude that the one-dimensional AFM
order starts to form at the temperature of 220\,K and that the corresponding N\'{e}el
temperature of this phase is near 140\,K. It remains unclear whether
nanoregions with a tripled unit cell, seen above 130\,K in the
electron diffraction, might be somehow related to a formation of short-range magnetic order below 220\,K. A three-dimensionally ordered AFM phase
appears at $\Tn \approx 59$\,K, where the magnetic susceptibility exhibits a minimum (Fig~\ref{fig:magnetization}).

\section{Conclusion}

Our complex structural, SHG, dielectric and lattice vibration studies of
  2H-BaMnO$_3$ ceramics revealed features related to the structural phase
  transition at $\Tc=130$\,K. Above \Tc, the phase transition is  driven by a
  soft phonon from the BZ boundary with a wavevector $q =
  (\frac{1}{3},\frac{1}{3},0)$. Below \Tc, BZ folding occurs and the soft mode
  becomes simultaneously IR- and Raman-active. On
  cooling, its frequency increases according to the Cochran law. The
  structure of the low-temperature phase
  was unambiguously determined as $P6_{3}cm$ with $Z=6$, but the symmetry of the
  high-temperature phase, in spite of our efforts, remains unclear.
 The electron and XRD data can be
  interpreted as corresponding either to a non-polar ($P6_{3}/mmc$)
  or a polar ($P6_{3}mc$) space group
  with $Z=2$. The weak SHG signal detected above \Tc{} does not
  represent an unambiguous
  evidence about a polar structure above \Tc, as it may be due
  to the sample surface.

The existence of a FE polarization was proved below \Tc,
but we were not able to determine it in the high-temperature
phase, owing to a leakage current in the
  pyrocurrent measurements above \Tc. Three observed modes
    simultaneously active in both IR and
  Raman spectra
  seem to support the
$P6_{3}mc$ space group (phonons in the $P6_{3}/mmc$ symmetry should have different
activities in IR and Raman spectra), but this evidence is rather weak.
First, only three out of seven simultaneously active phonons allowed by
factor-group analysis were observed. Second, sixteen modes were detected
  in the IR spectra, far more than the seven allowed in the $P6_{3}mc$ structure. The excess modes can be defect-induced or due to nanoclusters
of a polar $P6_{3}cm$ phase present above \Tc. The latter possibility was partially
supported by electron diffraction revealing the presence of $P6_{3}cm$ clusters at
least 40\,K above \Tc. These clusters may also explain the
observation of the three
phonons active simultaneously in IR and Raman spectra.

The Landau theory of an improper FE phase transition
 in 2H-BaMnO$_3$ from the $P6_{3}/mmc$ to the $P6_{3}cm$
symmetry was published by Varignon and Ghosez.\cite{Varignon13} We
presented an alternative Landau theory of a possible
nonferroic phase transition between two polar phases and
derived the phase diagram in Fig.~\ref{fig:phasdiag}. In both cases the phase
transition is driven by the observed BZ-boundary phonon with $q =
(\frac{1}{3},\frac{1}{3},0)$.

The temperature dependence of permittivity exhibits a change
in slope at \Tc. This
unusual behavior corresponds to a theoretical
$\varepsilon^{\prime}(T)$ dependence at
a second-order nonferroic phase transition,\cite{Toledano82} but our phase
transition between $P6_{3}mc$ and $P6_{3}cm$ space groups should be of the first order, where
some small jump is expected at \Tc.\cite{Toledano82} Also an improper FE
phase transition should exhibit a small jump or even a small peak in
$\varepsilon^{\prime}(T)$ at \Tc,\cite{Levanyuk74} which was not observed in our case.
Nevertheless, this characteristic change may be screened by the contribution of
FE domain-wall motions, which enhance the microwave and radio-frequency
permittivity below \Tc. A slowing down of domain-wall motions
was revealed below
60\,K, where the frequency-dependent decrease in $\varepsilon^{\prime}(T)$ was observed.
Slim S-shaped FE hysteresis loops were observed  below 130\,K.
Interestingly, the remnant polarization decreases below 60\,K and the loops
become pinched at 26\,K. On subsequent cooling the FE loops open again. We
cannot distinguish whether this is an intrinsic or an extrinsic effect;
to resolve this issue, further studies would be required.
For distinguishing clearly whether the high-temperature phase
  is polar or not, we
propose a SHG experiment using a high-quality single
crystal, where a higher laser
intensity could be applied. This would significantly enhance the
sensitivity of the SHG signal detection.

Our magnetic and EPR studies confirmed earlier observations of an AFM phase
transition at $\Tn \approx 59$\,K. Additionally, signatures of a
one-dimensional or short-range magnetic ordering were observed at least up to
220\,K. Again, new magnetic studies on a 2H-BaMnO$_3$ single crystal
should clarify the magnetic properties above \Tn.

\begin{acknowledgments}
This work was supported by the Czech Science Foundation, Projects Nos.
13-11473S, 14-14122P and 15-08389S, by the Czech Ministry of Education, Youth
and Sports, project LH15122, and by the program of Czech Research Infrastructures, project LM2011025.
\end{acknowledgments}


\end{document}